%% file: Anomalous_MC_arxiv.tex
\documentclass[aps,prb,amssymb,showpacs,twocolumn,floatfix]{revtex4-2}

\usepackage{hyperref}
\usepackage{lipsum}
\usepackage[version=3]{mhchem} 
\usepackage{miller}
\usepackage{graphicx}
\usepackage{amsmath}
\usepackage{color}
\usepackage[normalem]{ulem}
\usepackage{pdfpages}
\usepackage{etoolbox}
\usepackage{pdfpages}
\makeatletter
\patchcmd{\@outputpage@head}{\@ifx{\LS@rot\@undefined}{}{\LS@rot}}{}{}{}
\makeatother

\begin{document}
	\date{\today}
	
	\title[anomalous MC in oxide 2DES]{Dirac-like fermions anomalous magneto-transport in a spin-polarized oxide two-dimensional electron system}
	
	\author{Yu Chen\textsuperscript{1}} \email{yu.chen@spin.cnr.it}
	
	\author{Maria D'Antuono\textsuperscript{2,1}}
	
	\author{Mattia Trama\textsuperscript{3}}
	
	\author{Daniele Preziosi\textsuperscript{4}}
	
	\author{Benoit Jouault\textsuperscript{5}}
	
	\author{Frédéric Teppe\textsuperscript{5}}
	
	\author{Christophe Consejo\textsuperscript{5}}
	
	\author{Carmine A. Perroni\textsuperscript{2,1}}
	
	\author{Roberta Citro\textsuperscript{3,6}}
	
	\author{Daniela Stornaiuolo\textsuperscript{2,1}}\email{daniela.stornaiuolo@unina.it}
	
	\author{Marco Salluzzo\textsuperscript{1}} \email{marco.salluzzo@spin.cnr.it}
	
	\affiliation{\textsuperscript{1}{CNR-SPIN}, {Complesso Univ. Monte S. Angelo}, {Naples}, {I-80126}, {Italy}}
	
	\affiliation{\textsuperscript{2}{Department of Physics}, {University of Naples Federico II}, {Complesso Univ. Monte S. Angelo}, {Naples},{I-80126}, {Italy}}
	
	\affiliation{\textsuperscript{3}{Department of Physics}, {Universitá degli studi di Salerno}, {Via Giovanni Paolo II, 132}, {Salerno}, {I-84084}, {Italy}}
	
	\affiliation{\textsuperscript{4}{Institut de Physique et Chimie des Matériaux de Strasbourg}, {UMR 7504 CNRS and Université de Strasbourg}, {Strasbourg}, {F-67034}, {France}}
	
	\affiliation{\textsuperscript{5}{Laboratoire Charles Coulomb},{UMR 5221, CNRS, Université de Montpellier}, {Montpellier}, {F-34095}, {France}}
	
	\affiliation{\textsuperscript{6}{CNR-SPIN}, {Via Giovanni Paolo II, 132}, {Salerno}, {I-84084}, {Italy}}

\begin{abstract}
	
	{In two-dimensional electron systems (2DES) the breaking of the inversion, time-reversal and bulk crystal-field symmetries is interlaced with the effects of spin-orbit coupling (SOC) triggering exotic quantum phenomena. Here, we used epitaxial engineering to design and realize a 2DES characterized simultaneously by ferromagnetic order, large Rashba SOC and hexagonal band warping at the (111) interfaces between LaAlO$_{3}$, EuTiO$_{3}$ and SrTiO$_{3}$ insulators. The 2DES displays anomalous quantum corrections to the magneto-conductance driven by the time-reversal-symmetry  breaking occurring below the magnetic transition temperature. The results are explained by the emergence of a non-trivial Berry phase and competing weak anti-localization / weak localization  back-scattering  of Dirac-like fermions, mimicking the phenomenology of gapped topological insulators. These findings open perspectives for the engineering of novel spin-polarized functional 2DES holding promises in spin-orbitronics and topological electronics.}
	
\end{abstract}
\maketitle

\par
\begin{figure*}[ht!]
 \centering
		\includegraphics[width=145mm]{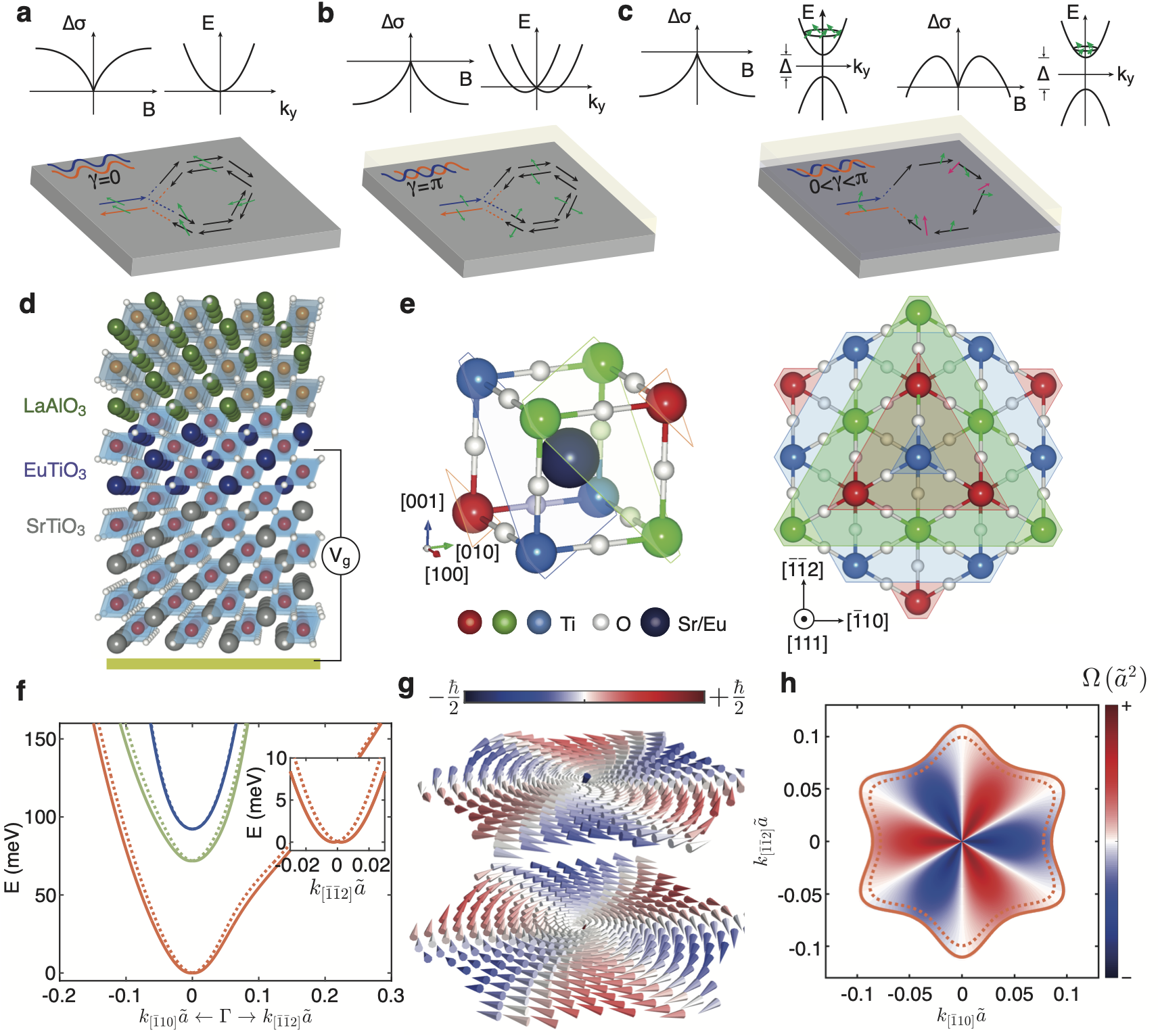}
		\caption{\label{fig1}\textbf{Schematics of WL, WAL, and WL/WAL back-scattering, and crystal, energy-bands, spin-texture and Berry curvature in (111) 2DES.} (a) Sketches of WL in a system without spin-momentum locking, e.g., disordered metals  with constructive interference of self-crossing loops and a zero Berry phase. (b) WAL in SOC systems, e.g., topological insulators and oxide interfaces, due to a destructive interference and a $\pi$ Berry phase. (c) Competing WL and WAL with a non-trivial Berry phase at low doping (right) in gapped Dirac systems, e.g., magnetically doped TIs \cite{Lu2011}. (d) Sketch of the LAO/ETO/STO heterostructure with a back-gate electrode. (e) The (Eu, Sr)TiO$_{3}$ lattice (left) and its projection (right) on the (111) interface/surface plane. Along the [111] interface/surface normal, the system consists of three non-equivalent layers of Ti atoms (red, green, blue), with the two perpendicular $[\bar{1}\bar{1}2]$  and $[\bar{1}10]$ crystallographic axes. (f) Electronic  band structure,  E(k) of (111) interfaces, showing bands with $a_{1g}$  (red) and $e_{g\pm}^{\pi}$ (green, blue) orbital character, together with their Rashba-like split counterparts (dark dotted lines). Throughout this letter, $k\tilde{a}$ are in units of $\pi$, where $\tilde{a} =\sqrt{2/3}a_0$  with the cubic STO lattice constant $a_0 = 3.905\AA$. (g) Spin arrangements of the lowest spin-split bands composed of $a_{1g}$ orbitals, exhibiting the emergence of an out-of-plane component and (h) corresponding Fermi contours shown as continuous/dotted lines and Berry curvature color map of a sub-band at E$_F$ = 60 meV.}
\end{figure*}

 Low dimensional electron systems can display exotic physical phenomena governed by the emergence of a non-trivial Berry curvature. The latter is linked to a band structure locally mimicking the dispersion relations of relativistic particles. The effects of these Dirac-like fermions in a 2DES are revealed when the chemical potential is tuned near the avoided crossing of Rashba-like SOC split bands. In three-dimensional (3D)-topological insulators (TIs) Dirac fermions drive intriguing quantum transport properties, such as the quantum anomalous Hall effect \cite{Haldane1988,YU2010}, topological phase transitions \cite{Bernevig2006} and changeovers from weak anti-localization (WAL) to weak localization (WL) quantum corrections to the magneto-conductance (MC) \cite{Lu2011,liu2012crossover, Lang2013}. In particular, the hexagonal band warping of most 3D-TIs can trigger a net out-of-plane spin polarization induced by in-plane magnetic fields, and the opening of scattering channels between multiple pairs of stationary points with opposite momenta \cite{Fu2009}. When a time-reversal-symmetry (TRS) breaking is introduced in such systems \cite{Lu2011,liu2012crossover,Lang2013}, the opening of a magnetic gap leads to the creation of an extra WL scattering channel, beside the WAL spin-orbit scattering associated to a $\pi$ Berry phase \cite{Berry1984,Shen2004}) (see Figures \ref{fig1}(a-c)). 
\par Among low dimensional electron systems, 2DES at the oxide interfaces are particularly appealing, as they can be opportunely engineered by epitaxy to host novel quantum phenomena, from unconventional superconductivity \cite{Caviglia2008,Stornaiuolo2016,Rout2017} to exotic magnetism and multiferroicity \cite{Brehin2023}. In the 2DES at the (001) interfaces between LaAlO$_{3}$ (LAO), EuTiO$_{3}$(ETO) and SrTiO$_{3}$ (STO) (LAO/ETO/STO), for instance, ferromagnetic correlations, due to the magnetic ordering of Eu\textsuperscript{2+} (4f\textsuperscript{7}) ions, coexist with a relatively large Rashba-like SOC and unconventional superconductivity \cite{Stornaiuolo2016,DiCapua2022}.
\par Recently, the study of (111) STO-based interfaces unveiled a large and unexpected second order bi-linear magneto-resistance and an anomalous planar Hall effect \cite{He2018, Lesne2023, Tuvia2024}. The results were interpreted as signatures of an out-of-plane spin-polarization and non-trivial Berry curvature nearby the Dirac-like point formed at crossing bands split, at finite momentum, by the Rashba-SOC. The phenomenon is triggered by a large external in-plane magnetic field and, in part, linked to the hexagonal band-warping of (111) heterostructures \cite{Lesne2023}.
\par Here, we show how epitaxial engineering of oxide heterostructures enables the realization of a rare example of oxide 2DES characterized by a non-trivial Berry phase ($\gamma \notin \{0, \pi \}$) without the application of an in-plane magnetic field. This non-trivial Berry phase gives rise to competing WL/WAL corrections to the MC due to Dirac-like fermions, with a phenomenology analogous to that of  gapped 3D-TIs \cite{Lu2011,liu2012crossover, Lang2013}.

\par The 2DES was realized by sequential epitaxial deposition of 3 unit cells (uc) delta-doping (111) ETO layer and 14 uc LAO on a Ti-terminated (111) STO single crystal (see Methods and Supplementary Fig. S1a). Magneto-transport properties were studied on Hall-bar devices patterned using photolithography and ion beam etching at low temperature \cite{DAntuono2022} (see  Methods). Transport characterizations \cite{Chen2022} confirm the formation of a 2DES with a sheet resistance temperature dependence similar to that of (001) and (111) LAO/STO heterostructures (Supplementary Fig. S1b). Tuning of the chemical potential was achieved by back-gating, as schematically shown in Fig. \ref{fig1}d. 
\par (111) STO-based oxide 2DES can be described as repetition of three inequivalent Ti-layers stacked along the [111] vector, i.e., perpendicular to the interface, arranged in a hexagonal lattice \cite{Xiao2011,Doennig2013,Khanna2019,Trama2022} (Fig. \ref{fig1}e). Inversion symmetry breaking gives rise to a $D_{3d}$ trigonal symmetry, resulting in hybridized $t_{2g}$ Ti-orbitals forming $a_{1g}$, $e_{g}^\pi$ states \cite{DeLuca2018}  (Fig. \ref{fig1}f and Supplementary Fig. S2(a,b)). The trigonal crystal field has several consequences on the electronic band structure of the 2DES. First, as shown by tight-binding calculations (Supplementary Note I), the Fermi surface of the lowest-energy bands exhibits a snowflake shape and a hexagonal band warping in a wide range of chemical potentials, affecting the electronic and transport properties \cite{Khanna2019}. Second, the hexagonal warping brings about an out-of-plane spin textures (see Fig. \ref{fig1}g and ref. \cite{He2018,Trama2022,Lesne2023}) and a Berry curvature with alternating positive and negative values around the Brillouin zone, as illustrated by numerical calculations for the lowest band with $a_{1g}$ orbital character (Fig. \ref{fig1}h). This Berry curvature leads to an overall $\pi$ Berry phase (Supplementary Note II). As a result, pure WAL corrections  to the low field MC are expected in these (111) systems that do not have a TRS breaking \cite{Lesne2023,Rout2017, Tuvia2024}.
\begin{figure*}[ht!]
\centering
    \includegraphics[width=126mm]{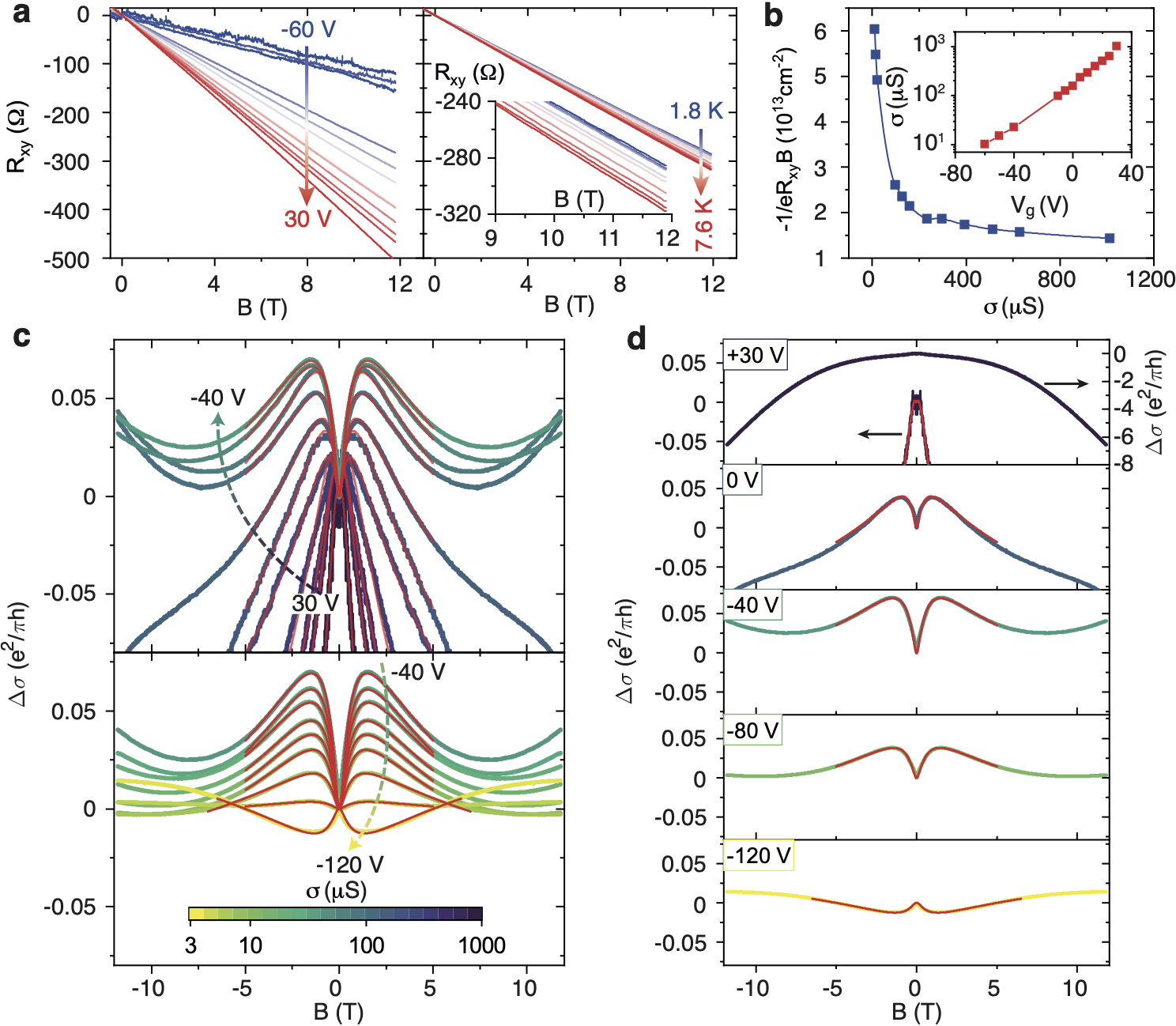}
    \caption{\label{fig2}\textbf{Magneto-transport properties of the (111) LAO/ETO/STO 2DES}. (a) 
    Gate voltage (at 1.8 K) and temperature (at V$_g$ = -10 V) dependence of the Hall effect on a Hall-bar oriented along the $[\bar{1}10]$  crystallographic direction. The inset is a zoomed-in view. (b) Inverse Hall coefficient versus the sheet conductance, tuned by the gate voltage as shown in the inset. (c) MC  as function of gate voltages in the range +30 V to -40 V (upper panel) and -40 V to -120 V (lower panel). MC data in full voltage range is also shown in Supplementary Fig. S4. The color-code of each MC data corresponds to the sheet conductance shown in the colorbar ($\sigma \approx 1\:mS - 2.9\:\mu S$). Red lines are the best fits using Equation (\ref{TI}). (d) Extracted MC data from the data set and their fits, at 30, 0, -40, -80 and -120 V, to highlight the evolution of the peak-shoulder feature as function of the gate voltage. The MC at +30 V in the full range is displayed in the right axis.}
\end{figure*}
\par On the other hand,  (111) LAO/ETO/STO  is characterized by ferromagnetically ordered Eu-ions, as confirmed by x-ray magnetic circular dichroism and SQUID magnetization measurements (see Supplementary Fig. S3 and ref. \cite{Chen2022}). The interaction between Eu-4f spins and Ti-3d moments induces a spin-polarization in the 2DES.  Moreover, the trigonal crystal field splitting in this system is three times larger than the one of the (111) LAO/STO system \cite{DeLuca2018}, enhancing the band splitting between the lowest $a_{1g}$ and $e_{g}^\pi$ derived bands. The enhanced splitting and the TRS-breaking driven by the ferromagnetic order are expected to induce novel quantum states, distinct from non-magnetic 2DES. 
\par To study the low temperature physics of (111) LAO/ETO/STO, we performed an extensive study of the gate-voltage dependent magneto-transport properties of the system (Fig. \ref{fig2}). In Fig.  \ref{fig2}a  we show the Hall effect data measured as function of the gate voltage and temperature. The Hall-effect resistivity is always linear, suggesting single-band transport of electrons belonging to the lowest $a_{1g}$ band. The inverse Hall coefficient decreases when increasing the gate voltage (Fig. \ref{fig2}b), which is opposite to the expected accumulation of electron-carriers observed for (001) oxide interfaces. This result is  a direct consequence of the band-warping, i.e., the non-parabolic nature of the $a_{1g}$ band. The snowflake Fermi contours of this band give rise to regions having positive and negative curvatures, thus opposite Fermi velocities, akin to (111) LAO/STO interfaces in the low doping regime (see ref. \cite{Khanna2019} and Supplementary Note III). 
\par In Fig. \ref{fig2}(c, d) we show the gate voltage V$_g$ dependence of the anomalous MC of (111) LAO/ETO/STO. At large negative gate voltage, i.e. -120 V, the MC is negative at low fields, similarly to what happens in the presence of WAL corrections (Fig. \ref{fig2}(d) bottom panel). Increasing the gate voltage, the shape changes drastically and the MC becomes positive with a cusp-like shape, followed by a peak and a shoulder. The peak to shoulder feature increases with V$_g$, reaches a maximum at  V$_g$=-40 V, and then decreases again until it almost (but not completely) disappears at +30 V (Fig. \ref{fig2}(d) upper panel). This behavior, in particular the positive cusp-like shape, is remarkably different from the positive parabolic WL or negative cusp-like WAL corrections observed in non-magnetic 2DES at the (001) and (111) LAO/STO \cite{Caviglia2010,Rout2017,Lesne2023} and LaTiO$_3$/STO interfaces \cite{Tuvia2024}. The latter were well explained by the models of Hikami-Larkin-Nagaoka, Iordanskii-Lyanda-Geller-Pikus, or Maekaewa-Fukuyama \cite{Hikami1980,IordanskiiS.V.Lyanda-GellerYu.B.1994,Maekawa1981}. On the other hand, the MC data of (111)LAO/ETO/STO (Fig. \ref{fig2}(c,d)) are reminiscent of the phenomenology observed in the magneto-transport of magnetically doped 3D-TIs \cite{Lang2013}, where it was attributed to competing WAL and WL channels induced by the opening of a magnetic gap $\Delta$ (see Figure \ref{fig1}(c)), and well reproduced by the formula derived in ref. \cite{Lu2011}: 
\begin{equation}\label{TI}
\delta \sigma(B)=\sum_{i=0,1} \frac{\alpha_i e^2}{\pi h}\left[\Psi\left(\frac{\ell_B^2}{\ell_{\phi i}^2}+\frac{1}{2}\right)-\ln \left(\frac{\ell_B^2}{\ell_{\phi i}^2}\right)\right],
\end{equation}
where $\Psi$ is the Digamma function, $\ell_B^2 = {\hbar}/{4eB}$ and $\ell_{\phi i}$ are the effective phase coherence lengths. The pre-factors $\alpha_0$ and $\alpha_1$, with opposite signs, represent the weights of the WAL and WL channels contributions. 
As shown in Fig. \ref{fig2}(c), where the fitting-curves are displayed as red-lines, Equation (\ref{TI}) captures well the data in the full range of gate voltages. As for 3D-TIs \cite{Lu2011}, both $\alpha_i$ and $\ell_{\phi i}$ are expected to be a function of the ratio between the magnetic gap and chemical potential  $\Delta$/E$_{F}$ (see Supplementary Fig. S5). The good agreement between the data and Equation (\ref{TI}) shown in Fig. \ref{fig2}(c,d) indicates that the low temperature anomalous MC is explained by two competing WL and WAL scattering channels related to the magnetic gap $\Delta$ opening, each contributing to the MC quantum corrections.

\begin{figure*}[ht!]
\centering
\includegraphics[width=1\linewidth]{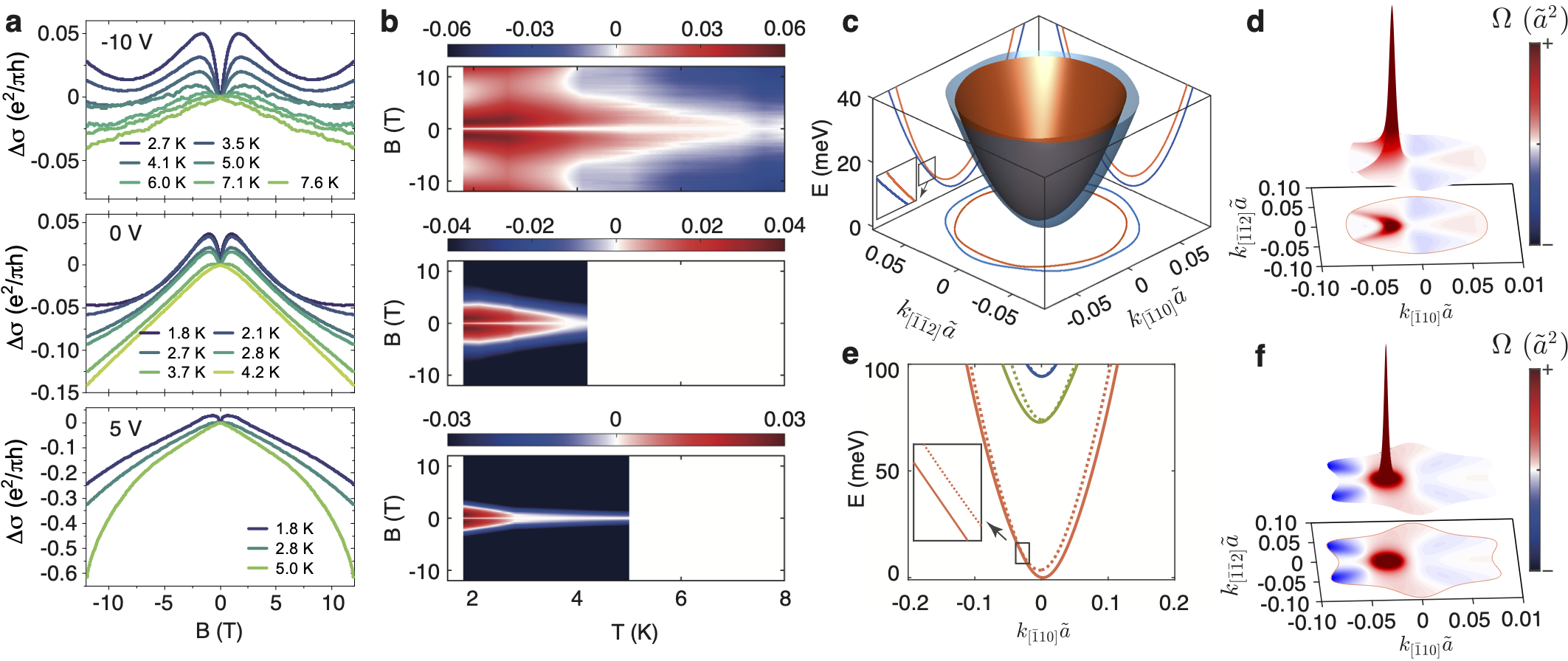}
\caption{\label{fig3} \textbf{Temperature dependence of MC, calculated band structures and Berry curvatures in a 2DES with in plane ferromagnetic correlations.} (a) MC data as function of the temperature at -10 V, 0 V, 5 V gate voltages, and (b) corresponding color-maps. The anomalous peak to shoulder feature disappears at slightly different temperatures depending on the gate voltage. (c) Energy bands in the presence of a planar exchange field with the direction along  $[\bar{1}\bar{1}2]$. Lowest Rashba-spin split energy bands and (d) Berry curvature at E$_F$ = 60 meV of a spin-split band analytically evaluated from the minimal model described in the main text. The calculations show the emergence of a hot-spot, i.e. a non-trivial Berry phase, near the Dirac-like point. Using tight-binding Hamiltonian (beyond the minimal model) band structure calculations of E(k)  along $[\bar{1}10]$ (e), and non-trivial Berry curvature (f). The avoided crossing of the bands and the opening of a magnetic gap are magnified in the insets of panels (c,e).}
\end{figure*}

\par In this framework, a purely WAL MC is predicted upon closing of the magnetic gap, which should take place approaching the FM transition of the system. In order to verify the link between the anomalous MC and the magnetism of the 2DES, in Fig. \ref{fig3}(a,b) we show the peak-shoulder feature evolution as a function of the temperature at different gate voltages. By increasing the temperature, the MC-shape changes completely within few Kelvins and a negative MC, i.e. WAL, is recovered at temperatures in the range between 5 and 8 K, compatible with the fading away of magnetic correlations induced by the Eu$^{2+}$ magnetic order (T$_c$ between 5 and 10 K) \cite{Stornaiuolo2016,Stornaiuolo2018}. Thus the experimental results show that the anomalous MC is linked to the magnetic gap, that is,  the band topology character. 
\par As mentioned above, in magnetically doped 3D-TIs this band topology gives rise to a non-trivial Berry phase (see Fig. \ref{fig1}c and ref. \cite{Lu2011}). Now we introduce a minimal model including Rashba-split bands derived from the lowest $a_{1g}$ orbital, a hexagonal warping term, and a magnetic exchange interaction related to the magnetic correlations in the 2DES. It is worth noting that within this minimal model, the hexagonal warping approximates well the band-structure of (111) 2DES only at low values of the momentum (Supplementary Fig. S2a). However, it has the advantage to provide analytical expressions for most of the quantities, which helps in understanding the main consequences of the interplay between the Rashba-SOC, the TRS-breaking, and the breaking of the bulk ($D_{4h}$) crystal-field symmetry into the $D_{3d}$ symmetry.  \par Magnetic correlations in the 2DES are included by adding an in-plane ferromagnetic magnetization of the form $H_{FM}  =  {\bf h}_{\vert\vert}\cdot\vec{\sigma} = \epsilon_{\vert\vert} (\sigma_x \cos\psi + \sigma_y \sin\psi )$ (Supplementary Note IV). Here $\psi$ is the angle between the magnetization direction and the in-plane axis $\hat{x}$  ($[\bar{1}10]$), and ${\bf h}_{\vert\vert}$ is the effective parallel field associated to the in-plane magnetic exchange $\epsilon_{\vert\vert}$.  The energy band dispersion assumes the form $E_{\pm} = \hbar^2\mathbf{k}^2/2m^* \pm \sqrt{\alpha^2 k^6 \cos^2(3\phi) + \lambda^2\tilde{k}^2} $, where tan($\phi$)=$k_y/k_x$, $\tilde{k}^2 = (k_x - \epsilon_{\vert\vert}\sin\psi/\lambda )^2 + (k_y + \epsilon_{\vert\vert} \cos\psi/\lambda )^2$, $m^*$ is effective mass, $\lambda$ is Rashba-SOC strength, and $\alpha$ is the warping parameter (Supplementary Note II). In the presence of the internal in-plane magnetic exchange, a magnetic gap $\Delta = 2\alpha (\epsilon_{\vert\vert}/\lambda)^3 \sin (3\psi)$ opens up because the warping term couples the in-plane magnetization to the out-of-plane spin (Fig. \ref{fig3}(c,f)), except for  $\psi = 2 \pi n/6$, i.e. along $[\bar{1}10]$ (Supplementary Note IV and Supplementary Fig. S6). The Dirac-like point is then shifted from the center of Brillouin zone to the $\bf{k^*} = \hat{z}\times {\bf{h}}_{\vert\vert}/\lambda$ position, where $\hat{z}$ is the unit vector along the [111] direction, leading to a non-trivial Berry curvature (Fig. \ref{fig3}d) and a hot-spot at the avoided crossing between the two Rashba-SOC spin-split bands \cite{Lesne2023}. 
\par In order to confirm the overall picture provided by the minimal model, we performed calculations  also using the full Hamiltonian tight binding approach, which better reproduces the snowflake shape of the Fermi surfaces. As shown in Fig. \ref{fig3}(e,f), the full Hamiltonian tight binding calculations confirm the emergence of a non-trivial Berry curvature and give an even stronger evidence of a hot-spot at the avoided crossing between the two Rashba-SOC spin-split bands (related spin textures shown in Supplementary Fig. S5(b-d)).
\par The evolution with the gate-voltage of the anomalous MC-data in the (111) LAO/ETO/STO 2DES can be, then, directly related to the tuning of the chemical potential: a maximum of the peak-shoulder feature associated  to the competing WL and WAL terms is attained when the chemical potential approaches the hot-spot, while departing from it drives the gradual reduction. Thus, the MC is strongly affected by the back-scattering of the Dirac-like fermions, experiencing a non-trivial Berry phase along self-crossing loops. This behavior, already established in gapped 3D-TIs \cite{Lu2011, liu2012crossover,Lang2013}, was rarely found in other 2DES, since it requires the simultaneous presence of a large Rashba-like SOC, a sizable gap $\Delta$ due to TRS-breaking, and very low chemical potential \cite{Lu2011}.
\par Our work establishes a method to create oxide 2DES characterized by transport properties mimicking those of systems hosting Dirac fermions, as in gapped 3D-TIs. The simultaneous presence of Rashba-SOC, magnetic correlations, and the hexagonal symmetry of the system with enhanced trigonal crystal field splitting, allows the tuning of the 2DES chemical potential near a Dirac-like point generated by the spin-split lowest energy bands. This triggers Berry-curvature hot-spot and topological charges in the 2DES without the need of external planar magnetic field, which are on the other hand needed in the case of non-magnetic 2DES, e.g., in (111) LAO/STO and LaTiO$_3$/STO \cite{Lesne2023, Tuvia2024}. 
Similar approaches can be envisaged in the case of other novel interfaces, including 2D-atomic non-magnetic/magnetic bi-layers and other oxide 2DES, alike superconducting EuO/KTaO$_{3}$ (111) based-2DES \cite{Liu2021}, opening a vast space for exploration at the intersection between topology and correlations of broad interest in the field of spin-orbitronics and topological electronics. 

\section*{Acknowledgments}
 This project has received funding from the European Union's Horizon Europe research and innovation programme project IQARO, under grant agreement n. 101115190, by the Ministry of University and Research project PRIN 2022 SONATA, No. P2022SB73K, funded by the EU - Next Generation EU, and PRIN 2022 STIMO, No. 2022TWZ9NR. M. S. and Y. C. acknowledge financial support from PNRR MUR project PE0000023-NQSTI. M.T. acknowledges financial support from "Fondazione Angelo Della Riccia”. DP thanks the French National Research Agency (ANR) through the ANR-JCJC FOXIES ANR-21-CE08-0021.  

\input{Anomalous_MC_111LAOETOSTO.bbl}

\section*{Methods}\label{methods}

\textbf{Sample growth}: Epitaxial LAO/ETO heterostructures were deposited on Ti-terminated STO(111) single crystals using pulsed laser deposition (PLD) assisted by reflection high energy electron diffraction (RHEED) \cite{Chen2022} at 720 $^\circ C$ and in a background oxygen pressure of  $7.5 \times 10^{-5}$ mbar. A KrF excimer laser with 248 nm  wavelength and 1 Hz repetition was focused on sintered Eu\textsubscript{2}Ti\textsubscript{2}O\textsubscript{7} target and on a crystalline LAO target at a fluence of 1.3 Jcm$^{-2}$. The sequential deposition of ETO (3 uc) and LAO (14 uc) films on Ti-terminated (111) STO single crystal was monitored by the oscillations of the specular RHEED intensity (Supplementary Fig. S1a).

\textbf{Device fabrication}: Different from the pattern technique of lithography - amorphous LAO deposition - liftoff - epitaxial film growth \cite{Chen2019a}, here
300 $\mu m$ $\times$ 100 $\mu m$ Hall bars were realized by a combination of optical lithography and cold ion milling  on the film LAO/ETO/STO in order to maintain the interface cleanliness as much as possible, following the procedure described in Ref. \cite{DAntuono2022}. 

\textbf{Magneto-transport measurements}: Magneto-transport measurements were performed in a He\textsuperscript{4} flow cryostat using standard lock-in amplifiers (f = 10 Hz and j$_{RMS}$ = 100 nA). The longitudinal and transverse voltages were measured simultaneously. The magnetic field was swept in the $\pm$ 12 T range. Data as function of the back-gate were acquired in a  +30 V to -120 V range. The leakage current across the (111) STO gate-oxide was below 1 nA. The sheet resistance saturates above +30 V. MC and Hall effect data were also acquired at different temperatures in the 1.8 K to  8 K range.

\textbf{XMCD and SQUID magnetization measurements}: Eu-M$_{4,5}$ edge XAS spectra were acquired at the Extreme beamline of the PSI-SLS and at the ID32 ESRF synchrotron facilities using the total electron yield (TEY) method (see ref. \cite{Chen2022}). The magnetic-field-dependent magnetization loops were obtained by measuring, at each field, the difference between the TEY intensities at the M$_{5}$-Eu edge peaks obtained with two different helicities (combination of polarization and field direction), and normalized to the background acquired at an energy below the absorption edge. SQUID data were collected by using a Quantum Design MPMS3 magnetometer at the Université de Strasbourg, IPCMS. The data were corrected for the STO diamagnetism subtracting the high field linear contribution.

\section*{Declarations}

\begin{itemize}
\item \textbf{Author contribution}\\
Conceptualization: YC, DS and MS. Sample preparation: YC, MS, MDA, DS. Transport measurements and analysis: YC, MDA, DS, BJ, DP, CC, FT and MS. SQUID experiment: DP. XMCD experiment: MS and DP. Theory modeling: YC, MT, CAP and RC.
Supervision: MS and DS. Writing: YC and MS. Writing-review and editing: all the authors participated in the editing of the paper.

\item \textbf{Data availability} 
Data needed to evaluate the conclusions of this paper are present in the Letter and its Supplementary Information. Additional data are available from the corresponding authors upon request.
\item \textbf{Code availability} \\
The codes used to evaluate the conclusions of this paper are available from the corresponding authors upon request.  

\item \textbf{Competing interests} \\
The authors declare no competing interests.

\item \textbf{Current address} Maria D'Antuono is currently employed at Instituto de Microelectronica de Barcelona (IMB-CNM, CSIC), Campus UAB, 08193 Bellaterra, Spain. Mattia Trama is currently working at Institute for Theoretical Solid State Physics, IFW Dresden, Helmholtzstr. 20, 01069 Dresden, Germany.

\end{itemize}

\clearpage



\section*{Supplementary material}

# Supplementary materials: Dirac-like fermions anomalous magneto-transport in a spin-polarized oxide two-dimensional electron system

Yu Chen[1*], Maria D'Antuono[2,1], Mattia Trama[3], Daniele Preziosi[4], Benoit Jouault[5], Frédéric Teppe[5], Christophe Consejo[5], Carmine A. Perroni[2,1], Roberta Citro[3,6], Daniela Stornaiuolo[2,1*], and Marco Salluzzo[1*]

[1*]CNR-SPIN, Complesso Univ. Monte S. Angelo, Naples, I-80126, Italy.
[2]Department of Physics, University of Naples Federico II, Complesso Univ. Monte S. Angelo, Naples, I-80126, Italy.
[3]Department of Physics, Universitá degli studi di Salerno, Via Giovanni Paolo II, 132, Salerno, I-84084, Italy.
[4]Institut de Physique et Chimie des Matériaux de Strasbourg, UMR 7504 CNRS and Université de Strasbourg, Strasbourg, F-67034, France.
[5]Laboratoire Charles Coulomb, UMR 5221, CNRS, Université de Montpellier, Montpellier, F-34095, France.
[6]CNR-SPIN, Via Giovanni Paolo II, 132, Salerno, I-84084, Italy.

*Corresponding author(s). E-mail(s): yu.chen@spin.cnr.it; daniela.stornaiuolo@unina.it; marco.salluzzo@spin.cnr.it;

## Contents

# 1 Supplementary Note I: Tight-binding model

The system can be theoretically described by a tight-binding model [1–3], (see Fig. 1f and Fig. 3(e,f) in the main text, and Fig. S2(a,b)), within a $\{|\uparrow\rangle,|\downarrow\rangle\} \otimes \{|Ti_1\rangle,|Ti_2\rangle,|Ti_3\rangle\} \otimes \{|d_{yz}\rangle,|d_{zx}\rangle,|d_{xy}\rangle\}$ basis, which is equivalent to the $\{|a_{1g}\rangle,|e_{g+}^{\pi}\rangle,|e_{g-}^{\pi}\rangle\} \otimes \{|Ti_1\rangle,|Ti_2\rangle,|Ti_3\rangle\} \otimes \{|\uparrow\rangle,|\downarrow\rangle\}$ basis of Ref. [2]. These two representations can be inter-converted by

$$\begin{pmatrix} |a_{1g}\rangle \\ |e_{g+}^{\pi}\rangle \\ |e_{g-}^{\pi}\rangle \end{pmatrix} = \frac{1}{\sqrt{3}} \begin{pmatrix} 1 & 1 & 1 \\ 1 & \omega & \omega^2 \\ 1 & \omega^{-1} & \omega^{-2} \end{pmatrix} \begin{pmatrix} |d_{xy}\rangle \\ |d_{yz}\rangle \\ |d_{zx}\rangle \end{pmatrix}, \quad (1)$$

where $\omega = e^{i\frac{2\pi}{3}}$. Following the treatment in Ref. [3], we write the Hamiltonian as:

$$H = H_0(\vec{K}) + H_{SO} + H_{cf} + H_R(\vec{K}), \quad (2)$$

where $\vec{K} = (K_1, K_2)$ is the quasi in-plane momentum, $K_1 = -\frac{\sqrt{3}}{2}\vec{K}_x + \frac{3}{2}\vec{K}_y$ and $K_2 = \frac{\sqrt{3}}{2}\vec{K}_x + \frac{3}{2}\vec{K}_y$. Here $\vec{K}_x$ is the momentum along the $[\bar{1}10]$ direction and $\vec{K}_y$ is the momentum along the $[\bar{1}\bar{1}2]$ direction. The in-plane lattice constant is $\tilde{a} = \sqrt{2/3}a_0$, where the cubic STO lattice constant $a_0 = 0.3905\,nm$. The hopping terms $H_0(\vec{K})$ are given by:

$$H_0(\vec{K}) = \sum_{\vec{k}} \sum_{\mu,\alpha\beta,\sigma} t_\mu^{\alpha\beta}\left(\vec{K}\right) d^\dagger_{\mu\alpha\sigma,\vec{K}} d_{\mu\beta\sigma,\vec{K}}, \quad (3)$$

where $\mu$ labels the $|d_{yz}\rangle, |d_{zx}\rangle, |d_{xy}\rangle$ orbitals, $\sigma$ the spin and $\alpha, \beta$ the Ti layers. The transfer matrix $t_\mu^{\alpha\beta}$ can be written in this basis as $\{|\text{Ti}_1\rangle,|\text{Ti}_2\rangle,|\text{Ti}_3\rangle\} \otimes \{|d_{yz}\rangle,|d_{zx}\rangle,|d_{xy}\rangle\}$:

$$t_\mu^{\alpha\beta}(\vec{K}) = \begin{pmatrix} A(\vec{K}) & B(\vec{K}) & \mathcal{O}_3 \\ B^*(\vec{K}) & A(\vec{K}) & B(\vec{K}) \\ \mathcal{O}_3 & B^*(\vec{K}) & A(\vec{K}) \end{pmatrix}, \quad (4)$$

The diagonal matrices are $A(\vec{K}) = \text{diag}(\tilde{\epsilon}_{yz}, \tilde{\epsilon}_{zx}, \tilde{\epsilon}_{xy})$ and $B(\vec{K}) = \text{diag}(\epsilon_{yz}, \epsilon_{zx}, \epsilon_{xy})$, where the next-to-nearest-neighbor (NNN) hopping elements are $\tilde{\epsilon}_{yz} = -2t_{\sigma''}\cos(K_1)$, $\tilde{\epsilon}_{zx} = -2t_{\sigma''}\cos(K_2)$, $\tilde{\epsilon}_{xy} = -2t_{\sigma''}\cos(K_2 - K_1)$ and the nearest-neighbor (NN) hopping elements are $\epsilon_{yz} = -t_\pi[1 + \exp(-iK_1)] - t_{\delta'}\exp(-iK_2)$, $\epsilon_{zx} = -t_\pi[1 + \exp(-iK_2)] - t_{\delta'}\exp(-iK_1)$, $\epsilon_{xy} = -t_\pi[\exp(-iK_1) + \exp(-iK_2)] - t_{\delta'}$. The NN hopping parameters are $t_\pi = 1.2\,eV$, $t_{\delta'} = 0.12\,eV$ and the NNN parameter is $t_{\sigma''} = 0.4\,eV$.

The single-particle Hamiltonian includes the trigonal crystal-field $\Delta_{cf} = -50\,meV$, which lifts the degeneracy between $|a_{1g}\rangle$ and $|e^{\pi}_{g\pm}\rangle$ orbitals at $\Gamma$ point:

$$H_{cf} = \frac{\Delta_{cf}}{2} \sum_{\vec{k}} \sum_{\mu \neq \nu, \alpha, \sigma} d^{\dagger}_{\mu\alpha\sigma,\vec{K}} d_{\nu\alpha\sigma,\vec{K}}. \tag{5}$$

In addition, the on-site atomic spin-orbit coupling $H_{SO} = \Delta_{SO} \vec{L} \cdot \vec{s}$ mixes the three orbitals $a_{1g}$, $e^{\pi}_{g+}$, $e^{\pi}_{g-}$ and splits the degeneracy at $\Gamma$. The last term describes the Rashba effect $H_R(\vec{K})$ which originates from the reflection symmetry breaking due to an effective field $\vec{E}$ along the [111] direction. Following the perturbation treatment in Ref. [3], we rewrite $H_R(\vec{K})$ as $H_R(\vec{K}) = H_{V_c} + H_{V_h}(\vec{K})$ in the $\{|\text{Ti}_1\rangle, |\text{Ti}_2\rangle, |\text{Ti}_3\rangle\} \otimes \{|d_{yz}\rangle, |d_{zx}\rangle, |d_{xy}\rangle\}$ basis, where the linear potential $V_c = 2\,eV$ is $H_{V_c}(\vec{K}) = V_c \cdot \text{diag}(2,1,0) \otimes \mathcal{I}_3$, thus, different in the three adjacent layers. A hopping term, odd in momentum, $H_{V_h}$ arises from the hybridization of $|d\rangle$ orbitals induced by $\vec{E}$ [3–5]. $H_{V_h}$ can be written as:

$$H_{V_h}(\vec{K}) = \Delta_{Rnn} R_{nn}(\vec{K}) + \Delta^{\pi}_{Rnnn} R^{\pi}_{nnn}(\vec{K}) + \Delta^{\sigma}_{Rnnn} R^{\sigma}_{nnn}(\vec{K}), \tag{6}$$

where the NN hopping terms between the adjacent layers are:

$$\tilde{R}_{nn} = \begin{pmatrix} 0 & e^{iK_1} - e^{iK_2} & 1 - e^{iK_2} \\ -e^{iK_2} + e^{iK_1} & 0 & 1 - e^{iK_1} \\ -1 + e^{iK_2} & -1 + e^{iK_1} & 0 \end{pmatrix}, \tag{7}$$

The full NN hopping matrix is then:

$$R_{nn} = \mathcal{I}_2 \otimes \begin{pmatrix} \mathcal{O}_3 & \tilde{R}_{nn} & \mathcal{O}_3 \\ -\tilde{R}^{*}_{nn} & \mathcal{O}_3 & \tilde{R}_{nn} \\ \mathcal{O}_3 & -\tilde{R}^{*}_{nn} & \mathcal{O}_3 \end{pmatrix}, \tag{8}$$

and the NNN hopping terms are written as:

$$\tilde{R}^{\pi/\sigma}_{nnn}(\vec{K}) = i \begin{pmatrix} 0 & -r^{\pi/\sigma}_1 & -r^{\pi/\sigma}_2 \\ r^{\pi/\sigma}_1 & 0 & -r^{\pi/\sigma}_3 \\ r^{\pi/\sigma}_2 & r^{\pi/\sigma}_3 & 0 \end{pmatrix}, \tag{9}$$

where $r^{\sigma}_1 = \sin(K_1) - \sin(K_2)$, $r^{\pi}_1 = \sin(K_1) + \sin(K_2 - K_1)$, $r^{\pi}_3 = -\sin(K_2) + \sin(K_2 - K_1)$, and $r^{\pi}_1 = r^{\sigma}_1 + 2\sin(K_2 - K_1)$, $r^{\pi}_2 = r^{\sigma}_2 - 2\sin(K_2)$, $r^{\pi}_3 = r^{\sigma}_3 + 2\sin(K_1)$. The full NNN hopping matrix is $\mathcal{I}_2 \otimes \mathcal{I}_3 \otimes (\tilde{R}^{\pi}_{nnn} + \tilde{R}^{\sigma}_{nnn})$. Finally, the prefactors are $\Delta_{Rnn} = \eta_p \frac{V_{pd\pi} 2^{7/8}}{\sqrt{15}}$, $\Delta^{\pi}_{Rnnn} = \eta_p \frac{2V_{pd\pi}}{\sqrt{15}}$ and $\Delta^{\sigma}_{Rnnn} = \eta_p \frac{2\sqrt{3}V_{pd\sigma}}{\sqrt{15}}$, with the Slater-Koster parameters $V_{pd\pi} = 28\,meV$ and $V_{pd\sigma} = -0.065\,meV$ and $\eta \sim \sqrt{3}V_c/10a_0$ (Ref. [3]).

**Table 1** Fitting parameters from Eq. (13), and residual sum of squares $RSS = \sum [E_{fit}^0(k) - E_{a_{1g}}]^2$.

| cut-off E (meV) | $m^*$ ($m_0$) | $\alpha$ ($meV \cdot nm^3$) | $\lambda$ ($meV \cdot nm$) | RSS ($[meV]^2$) |
| --- | --- | --- | --- | --- |
| 70 | 0.39 | 8.7  | 5.0 | 1279.5 |
| 60 | 0.38 | 10.6 | 5.0 | 709.13 |
| 50 | 0.37 | 13.0 | 5.0 | 353.42 |
| 40 | 0.36 | 15.9 | 5.1 | 144.19 |
| 30 | 0.34 | 20.1 | 5.2 | 42.58 |
| 20 | 0.33 | 26.0 | 5.4 | 7.85 |
| 10 | 0.32 | 41.0 | 5.5 | 0.72 |

## 2 Supplementary Note II: Minimal Model and Berry Curvature, Berry Phase

Since the microscopic Hamiltonian involves many degrees of freedom and different competing terms, a full analytical treatment is quite difficult. We can use, instead, a phenomenological two-band model, which can capture the topological nature of the bands and the $C_{3v}$ symmetry. The minimal Hamiltonian we used is the following :

$$H = \hbar^2 \mathbf{k}^2/2m^* - \lambda(k_y \sigma_x - k_x \sigma_y) + h_w \sigma_z, \tag{10}$$

where we consider the warping effect by the term :

$$h_w = \frac{\alpha}{2}[(k_x + ik_y)^3 + (k_x - ik_y)^3] = \alpha k^3 \cos(3\phi), \tag{11}$$

with $\phi = \arctan(k_y/k_x)$. In absence of the exchange field $h_w \sigma_z$, the eigenvalues are given as:

$$E_\pm^0 = \hbar^2 \mathbf{k}^2/2m^* \pm \sqrt{\alpha^2 k^6 \cos^2 3\phi + \lambda^2 k^2}. \tag{12}$$

To fit the lowest energy dispersion $E_{a_{1g}}$ of the spin-split bands determined from the Hamiltonian (2), we combine the positive and negative momentum eigenvalues of (12) to redefine a fitting expression as:

$$E_{fit}^0(k) = \Theta(-k) E_-^0 + \Theta(k) E_+^0, \tag{13}$$

where $\Theta$ is Heaviside step function. The minimal model (10) matches well the tight binding calculation in low momentum (low energy) range (see Table (1) and Supplementary Fig. S7). For the following calculations, we use as parameters the averages of the values obtained from the fitting and shown in Table (1): the effective mass $m^* = 0.35 m_0$, the Rashba SOC strength $\lambda = 5.2 \, meV \cdot nm$ (also see Ref. [6]) and the strength of warping $\alpha = 19.3 \, meV \cdot nm^3$, in agreement with the value obtained using full Hamiltonian tight binding model in the low Energy range (Fig. S2 (a,c)). In (111) LAO/STO system, the warping factor used in Ref. [7] was $\alpha = 0.1 E_F/k_F^3$, comparable to the range of fitting parameters shown in the Table (1) and Supplementary Fig. S7b.

The Berry curvature (BC) can be evaluated as [7]:

$$\Omega^z_\pm(k,\phi) = \pm \frac{\lambda^2 \alpha k^3 \cos(3\phi)}{\left[\lambda^2 k^2 + \alpha^2 \left(k^3 \cos(3\phi)\right)^2\right]^{\frac{3}{2}}}. \tag{14}$$

The Berry curvature reflects the $C_{3v}$ symmetry of the Hamiltonian and assumes everywhere a finite value, except at the degeneracy point of the bands ($k = 0$) where the sum of the two vanishes. (Fig. S2 (c)) We can calculate the Berry phase for the area enclosed by a contour $C$ defined by the Fermi surface:

$$\begin{aligned}
\gamma_\pm(C) &= \oint_C d\mathbf{k} \cdot \langle u_\pm(\mathbf{k},s)|i\Delta_\mathbf{k}|u_\pm(\mathbf{k},s)\rangle \\
&= \pm \int_0^{2\pi} d\phi \int_{0+\epsilon}^{k_F(\phi)} dk \frac{\lambda^2 \alpha k^3 cos 3\phi}{\left[\lambda^2 k^2 + \alpha^2 \left(k^3 cos 3\phi\right)^2\right]^{\frac{3}{2}}} \\
&\quad + \oint_{C=\epsilon e^{i\phi}} d\mathbf{k} \cdot \langle u_\pm(\mathbf{k},s)|i\Delta_\mathbf{k}|u_\pm(\mathbf{k},s)\rangle,
\end{aligned} \tag{15}$$

where $u_\pm(\mathbf{k}, s)$ is a wave function relative to the eigenvalues $E_\pm$ defined in the main text in absence of the magnetic field and $k_F$ is the Fermi momentum. The first term of Eq. (15) is the integral of the BC using Stokes' theorem. The first term is zero because of the $C_{3v}$. Regarding the second term, if the warping effect is ignored i.e. the Hamiltonian (10) contains only the Rashba term, $\gamma_\pm$ gives rise to a $\pm\pi$ Berry phase [8, 9] for the two different bands.

Berry curvature also admits a general gauge invariant expression of the form

$$\Omega^n = i \sum_{n' \neq n} \frac{\langle n|\partial H/\partial k_x|n'\rangle \langle n'|\partial H/\partial k_y|n\rangle - (k_x \leftrightarrow k_y)}{(\varepsilon_n - \varepsilon_{n'})^2}, \tag{16}$$

where $n$ refers to the $n$-th eigenvalue. This expression can be used in order to compute the Berry curvature of the full microscopic model in Eq. (2). This will be done in the following by considering only a subset of eigenstates relative to the lowest doublet with a $a_{1g}$ character.

## 3 Supplementary Note III: Transverse conductance

The anomalous behavior of the transverse conductance can be explained by the strong anisotropy of the lowest doublet with a $a_{1g}$ character in the band structure.

Due to the strong orbital anisotropy, two channels with opposite Fermi velocities corresponding to the different curvature of the lobes of the Fermi surface arise. A hint of the explanation of this behavior can be found in Ref. [2] where different scattering times for different orbitals is invoked. The competition between these two channels in each band having substantial hexagonal warping is responsible for the decrease of the

Hall coefficient with the chemical potential. When other bands come into play, a full electron-like behavior is achieved.

In our LAO/ETO/STO the chemical potential is always below the bottom of the higher energy bands, which explain the monotonous decrease of the Hall coefficient as shown in Fig. 2b of the main text. This is due to the different and much larger trigonal crystal field splitting of the LAO/ETO/STO interface compared to LAO/STO. In order to understand this behavior, we introduce the expression of the longitudinal and Hall current from supplemental material of ref. [2], within the single time-scattering approximation, as

$$\sigma = -e^2\tau \sum_m \int_{BZ} \frac{d^2K}{(2\pi)^2} \partial_\mu f[\epsilon_m(\mathbf{K})] \frac{1}{2} \left([v_m^X(\mathbf{K})]^2 + [v_m^Y(\mathbf{K})]^2\right), \quad (17)$$

and

$$\sigma_{xy} = -e^3 B\tau^2 \sum_m \int_{BZ} \frac{d^2K}{(2\pi)^2} \left(\partial_\mu f[\epsilon_m(\mathbf{K})]\right) v_m^X(\mathbf{K}) \times$$
$$\times \left[v_m^Y(\mathbf{K})\partial_{Kx} - v_m^X(\mathbf{K})\partial_{Ky}\right] v_m^Y(\mathbf{K}), \quad (18)$$

where $m$ runs over the bands, $\epsilon_m(\mathbf{K})$ is the energy of the $m$th band at momentum $\mathbf{K}$, $f[\epsilon_m(\mathbf{K})]$ is the corresponding Fermi-Dirac distribution, $v_m^i(K) = \partial_{Ki}\epsilon_m(\mathbf{K})$ is the corresponding $i$th component of the group velocity. By defining the matrix

$$\hat{\sigma} = \begin{pmatrix} \sigma & \sigma_{xy} \\ -\sigma_{xy} & \sigma \end{pmatrix}, \quad (19)$$

the Hall resistance $R_{xy}$ is the off-diagonal element of $\hat{\sigma}_{xy}^{-1}$. These expressions can be used for the microscopic model in Eq. (2) and in particular for the warping Hamiltonian in Eq. (10) in order to verify that the non-monotonic behavior depends strongly on the anisotropy of the lobes of the Fermi surface. Qualitatively, the anisotropy of the Fermi surface, and the different group velocities along the Fermi contour give rise to terms with opposite sign which compete and give rise to the observed behaviour.

## 4 Supplementary Note IV: Magnetization Effect

Our LAO/ETO/STO system presents sizable in-plane ferromagnetic corrections induced by the magnetic moment of Eu$^{2+}$ 4f$^7$ ions [10–12]. The magnetism of the interface is corroborated by the magnetic field dependence of Eu-$M_5$ XMCD intensity and SQUID magnetometry, as shown in Fig. S3 as well as Ref. [12].

We consider the planar exchange field $H_M = \epsilon \sin\theta \sigma_z + \epsilon \cos\theta(\cos\psi\sigma_x + \sin\psi\sigma_y)$ in the minimal model and $H_M = \epsilon \sin\theta J_{[111]} + \epsilon \cos\theta(\cos\psi J_{[\bar{1}10]} + \sin\psi J_{[\bar{1}\bar{1}2]})$ where $J$ is the total angular momentum in the tight-binding model. Here the strength of the Zeeman coupling $\epsilon = 1.2g\ meV$ with the factor $g = 1.28$ in the minimal model and $g = 2$ in the tight-binding model, respectively, in order to have the same shift of the Dirac-like point in these two models (see Fig. S6). In the main text, only in-plane

magnetization $\epsilon_{||}$ is considered in the Hamiltonian. We can write the total Hamiltonian as:

$$H = \frac{\hbar^2 \mathbf{k}^2}{2m^*} + h_w \sigma_z - \lambda(\sigma_x k_y - \sigma_y k_x) + H_M, \tag{20}$$

The Berry curvature can be evaluated in the presence of planar exchange field:

$$\Omega^z_\pm(k, \phi) = \mp \frac{2\lambda^2 \alpha k^3 cos3\phi - 3\epsilon_{||} \alpha k^2 sin(2\phi + \psi)}{2\left[\alpha^2 k^6 cos^2 3\phi + \left(k\lambda sin\phi - \epsilon_{||} cos\phi\right)^2 + \left(k\lambda cos\phi + \epsilon_{||} sin\phi\right)^2\right]^{\frac{3}{2}}}. \tag{21}$$

The electronic band is shown in the Fig. 3c with minimal model and Fig. 3e in presence of the in-plane magnetization along $[\bar{1}\bar{1}2]$. Both depict the avoided crossing band along $[\bar{1}10]$, resulting in non-trivial Berry curvature with a hotspot (Fig. 3(d,f)).

## 5 Supplementary Note V: Fitting

The formula (1) in the main text can capture well the MC data as function of gate voltage and temperature. Here the MC data in a range from negative to positive magnetic field, e.g, from -5 T to 5 T, are used to fit with formula (1) in the main text. To avoid the divergence of the Digamma and logarithmic functions, the data at zero field is omitted. The characterized lengths $\ell_{\phi i}$ and prefactors $\alpha_i$ are extracted. $\alpha_0 = -0.5$, $\alpha_1 = 0$ stands for a pure WAL, while $\alpha_0 = 0$, $\alpha_1 = 0.5$ for a pure WL [13]. In order to study the competing WL and WAL, we set the fitting condition $-\alpha_0 + \alpha_1 = 0.5$. Using the fitting strategy outlined above, we can reduce the arbitrariness of fitting with four parameters.

# 6 Supplementary Figures

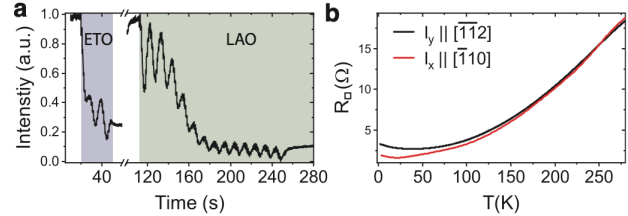

**Fig. S1 RHEED and transport measurement.** (a) RHEED oscillation and diffraction patterns of STO/ETO(3uc)/LAO(14uc) (b) Resistance as a function of temperature on the devices along $[\bar{1}10]$ and $[\bar{1}\bar{1}2]$ directions.

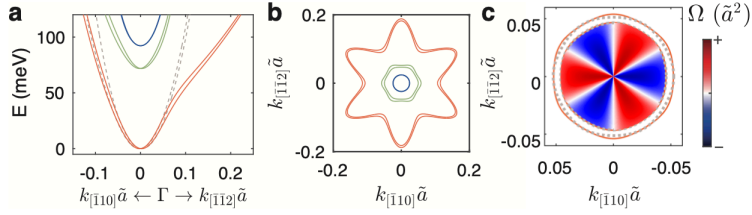

**Fig. S2 Theoretical calculation with the minimal and self-consistent full Hamiltonian tight binding model.** (a) Comparison between the electronic band structures calculated using the full Hamiltonian approach (continuous line) and the minimal model (dashed line). (b) Fermi contours at $E_F = 100$ meV. (c) Fermi contours of the $a_{1g}$ band at $E_F = 25$ meV obtained using the full Hamiltonian approach (red lines) and the minimal model (dashed lines), together with Berry curvature maps evaluated from Supplementary Equation (14).

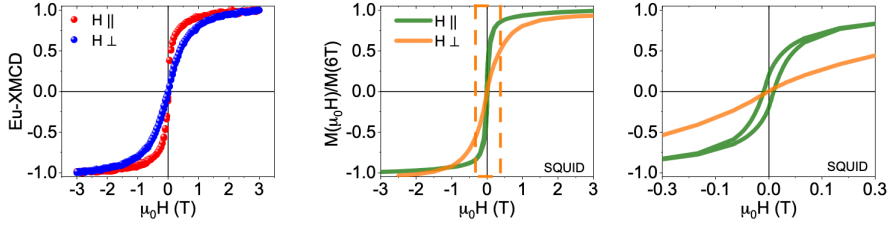

**Fig. S3 Magnetic characterizations.** (left-panel) $M_5$-edge Eu-XMCD intensity as function of the magnetic field perpendicular (blue) and parallel (red) to the interface. (middle-panel) Magnetic hysteresis loop acquired by SQUID (right-panel) SQUID data around zero magnetic field, to highlight the hysteresis of the low field in-plane magnetization.

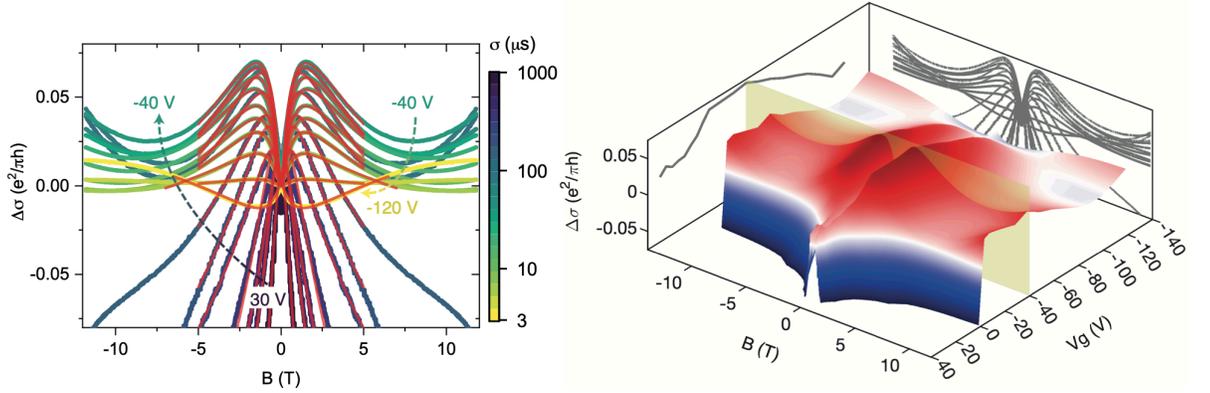

**Fig. S4 Gate Voltage dependence of the MC**. The MC peak to shoulder increases with gate voltages $V_g$, reaching a maximum at $V_g = -40\,V$. In the right panel, a 3D-view of the MC data, with a slice at -40 V.

9

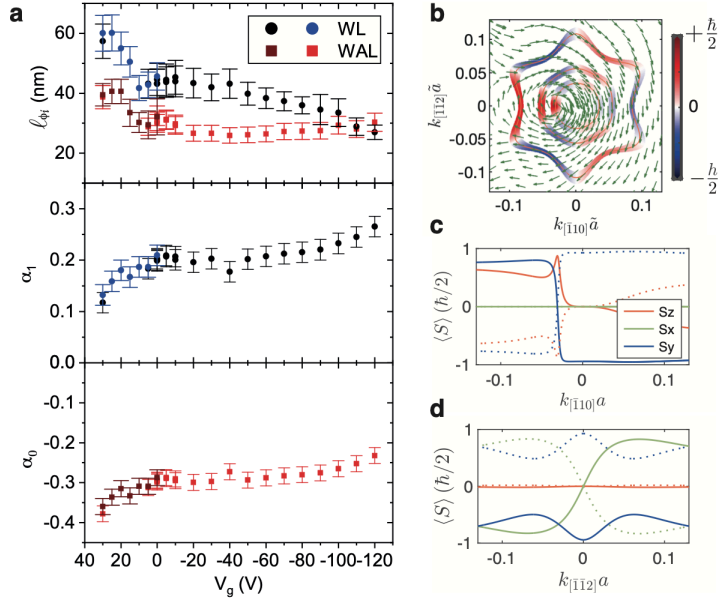

**Fig. S5 MC fitting parameters of MC data as function of voltage gate and spin alignment.** (a) Voltage gate dependence of MC data fitting parameters: characteristic scattering lengths $\ell_{\phi i}$ (upper), prefactors $\alpha_1$ representing for WL (middle) and $\alpha_0$ for WAL (bottom). The circles and squares denote the parameters of the WL-like contribution $\{\ell_{\phi 1}, \alpha_1\}$, and WAL-like ones $\{\ell_{\phi 0}, \alpha_0\}$, respectively. Two sets of data were measured on the same devices in different periods. (b) Spin-arrangement and contour map of the average out-of-plane spin moment, $\langle S_z \rangle$, of the single lowest band, in the presence of a planar exchange field along the $[\bar{1}\bar{1}2]$ direction at $E_F = 60$ meV (outer Fermi contour), 28 meV (middle Fermi contour), and 12 meV (inner Fermi contour), showing a hot-spot. Extracted profiles of average spin $\langle S \rangle$ along (c) $[\bar{1}10]$ and (d) $[\bar{1}\bar{1}2]$ exhibits a net out-of-plane spin moment (sum of $\langle S_z \rangle$) apart from the net $\langle S_y \rangle$ which is generated by the planar exchange field along the $[\bar{1}\bar{1}2]$ direction. Solid lines are for the bottom spin-band and dotted-lines are for the upper spin band calculated using the full-Hamiltonian approach.

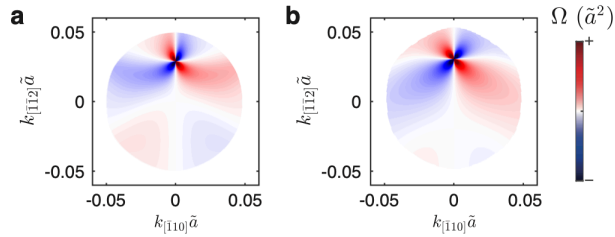

**Fig. S6 Berry curvature in the presence of a planar exchange field along $[\bar{1}10]$.** Calculation using minimal model (a) and full Hamiltonian tight-binding model (b) at $E_F = 30$ meV. In this case, the Berry phase is trivial due to the remaining mirror symmetry. These two models match well in low energy in spite of a slight discrepancy in the Fermi surface.

10

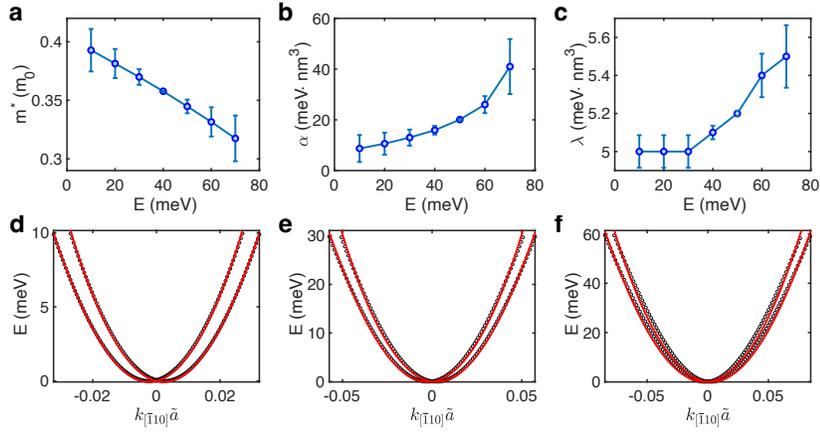

**Fig. S7 Fitting the lowest band dispersion of the minimal model.** (a-c) The fitting parameters including the effective mass $m^*$, Rashba SOC strength $\lambda$, the strength of warping $\alpha$. (d-f) The best fitting at the cut-off Energy 10, 30, 60 meV. The black circles denote the data calculated by tight binding model and the red lines denote fitting by minimal model.

11



\end{document}

%% file: Anomalous_MC_111LAOETOSTO.bbl
%